\documentclass[
showkeys,
preprintnumbers,amsmath,amssymb]{revtex4}

\usepackage{graphicx}
\usepackage{dcolumn}
\usepackage{bm}

\usepackage{graphicx}
\usepackage{epsfig}
\usepackage{color}

\begin{document}

\title{BAROTROPIC FRW COSMOLOGIES WITH CHIELLINI DAMPING}

\author{Haret C. Rosu}
\email{hcr@ipicyt.edu.mx}
\affiliation{IPICyT, Instituto Potosino de Investigacion Cientifica y Tecnologica,\\
Camino a la presa San Jos\'e 2055, Col. Lomas 4a Secci\'on, 78216 San Luis Potos\'{\i}, S.L.P., Mexico}

\author{Stefan C. Mancas }
\email{stefan.mancas@erau.edu}
\affiliation{Department of Mathematics, Embry-Riddle Aeronautical University,\\ Daytona Beach, FL. 32114-3900, U.S.A.}

\author{Pisin Chen}
\email{pisinchen@phys.ntu.edu.tw}
\affiliation{Leung Center for Cosmology and Particle Astrophysics (LeCosPA) and Department of Physics, National Taiwan University, Taipei 10617, Taiwan\\}

\begin{abstract} It is known that barotropic FRW equations written in the conformal time variable can be reduced to simple linear equations
for an exponential function involving the conformal Hubble rate.
Here, we show that an interesting class of barotropic universes can be obtained in the linear limit of a special type of  nonlinear {\em dissipative} Ermakov-Pinney equations with the nonlinear dissipation built from Chiellini's integrability condition. These cosmologies, which evolutionary are similar to the standard ones, correspond to barotropic fluids with adiabatic indices rescaled by a particular factor and have amplitudes of the scale factors inverse proportional to the adiabatic index.
\end{abstract}


\begin{center}
Phys. Lett. A 379 (2015) 882-887\\
{\tiny Issues 10-11}
\end{center}

\maketitle
%
\section{The FRW barotropic oscillator}

In conformal time $\eta$, the scale factors of the FRW barotropic universes, $a(\eta)$, assuming normalized-to-unit amplitude, have the following simple expressions
\begin{eqnarray}\label{ba1}
\begin{array}{lll}
a_{-}(\eta)&=\big[\sinh \bar\gamma (\eta -\eta_0)\big]^{\frac{1}{\bar{\gamma}}},   &\kappa =-1~,\\
a_{0}(\eta)&=(\eta-\eta_0)^{\frac{1}{\bar{\gamma}}}, & \kappa =0~,\\
a_{+}(\eta)&=\big[\cos \bar\gamma (\eta-\eta_0)\big]^{\frac{1}{\bar{\gamma}}},  &\kappa =+1~,\end{array}
\end{eqnarray}
where $\bar{\gamma}=3\gamma/2-1$ with $\gamma$ the adiabatic index, and $\eta_0$ an arbitrary constant. The case $\bar{\gamma}=0$ should be treated separately and does not enter the considerations in the following. The scale factors in (\ref{ba1}) correspond to
the three cases of the curvature index $\kappa$, i.e., $\kappa=-1$ for an open universe,
 $\kappa=0$ for a flat universe, and $\kappa=1$ for a closed universe, and
have entered textbooks since several decades \cite{wein,mis,lali}. They can be obtained by integrating the following second order nonlinear differential equation in conformal time $\eta$ ($^{\prime}=d/d\eta$)
\begin{equation}\label{conf1}
aa^{\prime\prime}+(\bar{\gamma}-1)a^{\prime 2}+\kappa \bar{\gamma}a^2=0~,
\end{equation}
which is obtained from the comoving Einstein-Friedmann dynamical equations after performing the change $dt=ad\eta$ from comoving to conformal time, and making usage of the barotropic equation of state $p=(\gamma-1)\rho$.

\medskip

Furthermore, by using ${\cal H}=a^{\prime}/a$, where ${\cal H}$ is the Hubble parameter in conformal time, one can transform (\ref{conf1}) to the Riccati equation
\begin{equation}\label{ricc}
{\cal H}^{^{\prime}}+\bar{\gamma}{\cal H}^{2}+\kappa\bar{\gamma} =0~,
\end{equation}
which for $\kappa=1$ corresponds to a harmonic oscillator of frequency $\bar{\gamma}$, while for $\kappa=-1$ corresponds to a hyperbolic oscillator.
The reduction of the barotropic FRW equations to these simple oscillator systems has been first obtained by Barrow \cite{b93,b86} and later the Riccati approach for cosmological barotropic fluids in conformal time has been pursued by several authors \cite{Far1,Far2,robar1,robar2,robar3,robar4}. Recently, the Riccati framework has been developed for scalar field FRW cosmologies by Harko {\em et al} \cite{Har-Ricc}, while El-Nabulsi obtained a Riccati equation for the `generalized time-dependent Hubble parameter' $H_g(t)=\Gamma(\alpha)t^{1-\alpha}H(t)$, where $\Gamma(\alpha)$ is the Euler Gamma function and $\alpha$ is an arbitrary real number. Notice that $\alpha=1$ corresponds to the standard FRW cosmology \cite{ElN1,ElN2}. In fact, by introducing the modified Hubble parameter in conformal time ${\cal H}_u(\eta)=u^{\prime}/\bar{\gamma}u$ \cite{Far1,Far2,robar1,robar2,robar3,robar4}, the Riccati equation (\ref{ricc}) can be reduced to the linear second order equation
\begin{equation}\label{w}
u\rq{}\rq{}+\kappa\bar{\gamma}^{2}u=0~,
\end{equation}
from which the set (\ref{ba1}) of scale factors are immediately obtained through $a(\eta)=[u(\eta)]^{1/\bar{\gamma}}$.
Although the FRW barotropic models look more as pedagogical examples, they are of considerable interest at the forefront of cosmology as several barotropic classes of fluids have been proposed in recent years as models for the enigmatic dark energy component of the universe \cite{ls09}.

\medskip

In this Letter, we will use the feature that the powers of order $\bar \gamma$ of the scale factors, i.e., the $u$ functions, are the solutions of the simple differential equation (\ref{w}), to make connections with another well-known equation in mathematical physics which in this way can be introduced in the barotropic cosmological context. This is the Ermakov-Pinney (EP) equation with  inverse cubic nonlinearity which has long been known to have profound connections with the linear equations of identical operatorial form, and because of this it has been considered as an example of `nonlinearity from linearity' \cite{KD}.
Previously, some EP equations have found their way in different cosmological setups \cite{ro,hl,pe,herring,daw}, but not in the barotropic FRW cases. Here, we will introduce non-standard barotropic cosmologies that are based on a modified form of equation (\ref{w}) obtained from the linear limit of the EP equations with an additional nonlinear dissipation term of the Chiellini type \cite{prev}.

\section{The FRW barotropic cosmologies based on Ermakov-Pinney solutions}

\subsection{The non-dissipative case}

\noindent The EP equations that corresponds to the cosmological barotropic oscillator (\ref{w}) are
\begin{equation}\label{a13}
v^{\prime \prime}+\kappa\bar{\gamma}^2v+{\rm k}v^{-3}=0~,
\end{equation}
where ${\rm k}$ is an arbitrary negative real constant which defines the strength of the inverse cubic nonlinearity. For ${\rm k}=0$, one recovers the linear equation (\ref{w}).
For arguments in the following, it is also convenient to write (\ref{a13}) in the form
\begin{equation}\label{a13b}
 v^{\prime \prime}
 +h(v)=0~, \qquad h(v)=\kappa \bar{\gamma}^{2}v+{\rm k} v^{-3}~.
\end{equation}

The particular solutions of the EP equation are given by (see also \cite{man})
\begin{eqnarray}\label{a14}
 \begin{array}{ll}
v_{-}(\eta;{\rm k})=\sqrt{-1+\Big(1-\frac{{\rm k}}{\bar{\gamma}^2}\Big)\,{\rm cosh}^2\bar\gamma(\eta-\eta_0)} ~, & \kappa =-1~,\\ 
\\
v_{0}(\eta;{\rm k})=\sqrt{(\eta-\eta_0)^2-{\rm k}}~, & \kappa=0~,\\
\\
v_{+}(\eta;{\rm k})=\sqrt{1-\Big(1+\frac{{\rm k}}{\bar\gamma^2}\Big)\sin^2\bar\gamma(\eta-\eta_0)}~, & \kappa=1~,
\end{array}
\end{eqnarray}
if one makes usage of the Pinney superposition formula \cite{P}
\begin{equation}\label{a11}
v(\eta;{\rm k})=\sqrt{u_1^2-\frac{{\rm k}u_2^2}{W^2}}~,
\end{equation}
where $W=\bar{\gamma}$ is the Wronskian of the two linearly independent solutions $u_1$ and $u_2$ of (\ref{w}).

\medskip

On the other hand, we can use the inverse relationship $u=f(v)$, as discussed by Steen \cite{steen1874} and Milne \cite{M30}, to make contact with the barotropic FRW cosmologies in conformal time
\begin{equation}\label{milne}
u(\eta)=\lim_{{\rm k}\rightarrow 0} v(\eta;{\rm k})\cos\left(\sqrt{{-\rm k}}\int^{\eta}\frac{d\eta}{v^2} +\varphi\right)~, \qquad \varphi\, - {\rm an}\,\, {\rm arbitrary}\,\, {\rm phase}~,
\end{equation}
where the integrals of the type $\int_{\eta_0}^{\eta}\frac{d\eta}{v^2}$ are known as Milne phases and are used in eigenvalue problems for Sturm-Liouville type differential equations \cite{yano4}.
Performing explicitly the limit, one finds that the nonlinear constant ${\rm k}$ occurs only as an additional factor for the known linear solutions used to construct the scale factors. In the next subsection, we will show that if one works with the special type of Chiellini-dissipative EP equations some interesting physical effects can be traced out.

\subsection{The Chiellini-dissipative case}

\noindent We introduce now the Chiellini dissipative EP equation as an equation having the same $h$ function as in (\ref{a13b}), but with an additional damping term
 \begin{equation}\label{v-heq}
 \tilde{v}^{\prime \prime}+g( \tilde{v})  \tilde{v}^{\prime}+h(\tilde{v})=0~,
 \end{equation}
 where the damping coefficient $g(\tilde{v})$ will be given by Chiellini's integrability condition for the Abel equation of the second kind which can be obtained from (\ref{v-heq}) by letting $\tilde{v}^\prime=z(\tilde{v}(\eta))$
\begin{equation}\label{12}
z\frac{dz}{d\tilde{v}}+gz+ h=0~.
\end{equation}
Then, in (\ref{12}) we use the inverse transformation $z=\frac 1 y$ which turns it into Abel's equation of the first kind
\begin{equation}\label{13}
\frac{dy}{d\tilde{v}}=g(\tilde{v})y^2+h(\tilde{v})y^3.
\end{equation}
 Equation (\ref{13}) is integrable if the dissipation function $g( \tilde{v})$ is obtained from $h( \tilde{v})$ by means of Chiellini's condition
\begin{equation}\label{Chiell}
\frac{d}{d \tilde{v}}\left(\frac{h( \tilde{v})}{g( \tilde{v})}\right)=pg( \tilde{v}),\qquad \qquad p, \, {\rm a\,\, real\,\, constant}.
\end{equation}
In the following, we will make use of dissipation functions $g$ that fulfill Chiellini's integrability condition that we will call Chiellini damping.

\medskip

The Chiellini-damped EP equation (\ref{v-heq}) has the interesting property that it can be turned into the nondissipative EP equation            
  \begin{equation}\label{v-meq}
 \tilde{v}^{\prime \prime}+\tilde{h}(\tilde{v})=0~,\qquad \tilde{h}(\tilde{v})=2h(\tilde{v})
 \end{equation}
having the $h$ function scaled up by a factor of two when the Chiellini condition (\ref{Chiell}) is satisfied for $p=-2$.
We further point out that this equivalence allows to deduce both (\ref{v-meq}) and (\ref{v-heq}) as equations of motion from the Lagrangian function
$$
L(q,q^{\prime})=\frac{q^{\prime 2}}{2}-\kappa\bar{\gamma}^2q^2+{\rm k}q^{-2}
$$
upon the identifications ${\tilde v}=q$ and ${\tilde v}^{\prime}=p$, where $q^{\prime}=p$. This Lagrangian, together with its corresponding Hamiltonian, are typical for an isotonic oscillator of potential $V(q)=\kappa\bar{\gamma}^2q^2-{\rm k}q^{-2}$ describing the Newtonian motion of a particle under the action of a linear force, and an additional inverse cube force with respect to the origin \cite{ran}.

\medskip

To prove that the Chiellini-dissipative EP equation is equivalent to a nondissipative EP equation with a scaled $h$, one needs to take
\begin{equation}\label{ve}
{\tilde v}\rq{}=\frac{h}{g}
\end{equation}
in (\ref{v-heq}). Next, let us differentiate (\ref{ve}) with respect to $\eta$ to obtain
\begin{equation}
{\tilde v}\rq{}\rq{}=\frac{d}{d \tilde{v}}\left(\frac{h( \tilde{v})}{g( \tilde{v})}\right)\tilde{v}\rq{}= \frac{h}{g}\frac{d}{d \tilde{v}}\left(\frac{h( \tilde{v})}{g( \tilde{v})}\right),
\end{equation}
and by equation (\ref{Chiell}) we obtain
\begin{equation}
{\tilde v}\rq{}\rq{}=ph(\tilde v)
\end{equation}
which is exactly (\ref{v-meq}) when $p=-2$. The remarkable feature of this result is that it allows us to find the dissipation $g(\tilde {v})$ of  (\ref{v-heq}) without knowing the solution $\tilde v$ as follows.

\medskip

Multiplying (\ref{v-meq}) by ${\tilde v}\rq{}$
\begin{equation}\label{bar3}
{\tilde v}\rq{}{\tilde v}\rq{}\rq{}+2h(\tilde{v})\tilde{v}\rq{}=0
\end{equation}
which also can be written as
\begin{equation}
\frac{d}{d \eta}({\tilde v}\rq{})^2+4h(\tilde v)\frac{d \tilde v}{d \eta}=0
\end{equation}
and by one integration leads to
\begin{equation}\label{bar5}
({\tilde v}\rq{})^{2}+4\int^ {\tilde v}  h({\tilde v})d{\tilde v}=c_1~.
\end{equation}
Thus:
\begin{equation}\label{bar6}
{\tilde v}\rq{}=\sqrt{c_1-4\int^{\tilde v}  h({\tilde v})d{\tilde v}}
\end{equation}
and now we use (\ref{ve}) to obtain
\begin{equation}\label{bar7}
g({\tilde v})=\frac{h({\tilde v})}{\sqrt{c_1-4\int^{\tilde v} h({\tilde v})d{\tilde v}}}~.
\end{equation}
From (\ref{bar6}) by one quadrature, we have
\begin{equation}\label{bar8}
\int^{\tilde{v}}\frac{d{\tilde v}}{\sqrt{c_1-4\int^{\tilde v}  h({\tilde v})d{\tilde v}}}=\eta-\eta_0~.
\end{equation}
If we define
\begin{equation}\label{bar9}
I_h({\tilde v})=\int^{\tilde{v}} \frac{d{\tilde v}}{\sqrt{c_1-4\int^{\tilde v} h({\tilde v})d{\tilde v}}}
\end{equation}
then the solution to  (\ref{v-meq}) is found from (\ref{bar8}) via the inversion
\begin{equation}\label{bar10}
{\tilde v}=I_h^{-1}(\eta-\eta_0),
\end{equation}
where $\eta_0$ depends on an initial condition.
Also, the nonlinear equation becomes linear with $g({\tilde v}),h({\tilde v})$ given by
\begin{equation}
g({\tilde v})=\tilde{g}(I_{h}^{-1}(\eta-\eta_0))~,\qquad  h({\tilde v})=\tilde{h}(I_{h}^{-1}(\eta-\eta_0))
\end{equation}
and with solution to the linear dissipative equation
\begin{equation}\label{bar11a}
v\rq{}\rq{}+\tilde{g}(\eta)v\rq{} +\tilde{h}(\eta)=0
\end{equation}
 given by (\ref{bar10}).

Using $h$ from (\ref{a13b}) together with (\ref{bar7})  leads to 
\begin{equation}\label{a19}
g(\tilde{v})=\frac{\kappa \bar \gamma^2\tilde{v}^2+{\rm k}\tilde{v}^{-2}}{\sqrt{-2 \kappa \bar \gamma^2 \tilde{v}^4+c_1 \tilde{v}^2+2{\rm k}}}~.
\end{equation}
Furthermore, from (\ref{bar8}) one has
\begin{equation}\label{a21}
\eta-\eta_0=\int^{\tilde v} \frac{\tilde{v}d\tilde{v}}{\sqrt{-2\kappa \bar \gamma^2\tilde{v}^4+c_1\tilde{v}^2+2{\rm k}}}~.
\end{equation}

By integration and inversion we then have the following general solutions of (\ref{v-heq}):
\begin{eqnarray}\label{aaatheta}
\begin{array}{lll}
\tilde{v}_{-1}(\eta;c_1,{\rm k})&=
\begin{array}{ll}
\frac{1}{2 |\bar \gamma|}\sqrt{-c_1+\sqrt{-\Delta^{-}_{\rm k}}\cosh \big(2 \sqrt 2\bar \gamma(\eta-\eta_0)\big)}~, \quad \Delta^{-}_{\rm k}<0~,   \kappa=-1~,
\\
\end{array} \\ 
\\
\tilde{v}_{0}(\eta;c_1,{\rm k})&=
\begin{array}{ll}
\sqrt{c_1 \eta^2-\frac{2{\rm k}}{c_1}} ~,   & \kappa=0~,
\end{array}\\
\\
\tilde{v}_{1}(\eta;c_1,{\rm k})&= \begin{array}{ll}\frac{1}{2 |\bar \gamma|}\sqrt{c_1+\sqrt{\Delta^{+}_{\rm k}}\sin \big(2 \sqrt 2\bar \gamma(\eta-\eta_0)\big)}\,
~,  \quad \Delta^{+}_{\rm k}>0~,   \kappa=1~
~,\\
\end{array} 
\end{array}
\end{eqnarray}
where $\Delta^{-}_{\rm k}=16{\rm k} \bar \gamma^2-c_1^2$, $\Delta^{+}_{\rm k}=16{\rm k} \bar \gamma^2+c_1^2$.
Examining this set of dissipative EP solutions, one can see that the differences with respect to the non-dissipative case
consist only in the presence of the new integration constant $c_1$ and the $\sqrt{2}$ scaling of $\bar{\gamma}$ which leads to a different frequency of the oscillatory ripples in the closed universe case. However, the reduction to the linear case by setting ${\rm k}=0$ does not lead this time to the standard barotropic cosmologies. This reduction will be discussed in the next subsection.

\medskip

\subsection{The reduced Chiellini-dissipative case}

\noindent At first glance, the solutions (\ref{aaatheta}) do not seem to offer anything appealing in cosmology. However, we will show now that taking the limit of the nonlinear coupling constant ${\rm k}=0$ in (\ref{v-heq}) does not lead to known results as this happens in the non-dissipative case. By taking this limit in the flat case, one obtains the linear equation as before. However, in the non-flat cases the reduced equation is still nonlinear because the Chiellini dissipation does not vanish in the limit
 \begin{equation}\label{v-hred}
 \tilde{u}^{\prime \prime}
 +g_{\kappa}( \tilde{u}) \tilde{u}^{\prime}
 +\kappa\bar\gamma^2 \tilde{u}=0~, \qquad g_\kappa( \tilde{u})=\frac{\kappa \bar\gamma^2 \tilde{ u}}{\sqrt{c_1-2\kappa\bar\gamma^2 \tilde{u}^2}}~.
 \end{equation}
This equation has the curious property that despite being nonlinear, it has in the case $\kappa=1$ the linear harmonic solutions
 \begin{eqnarray}\label{v-hred1}
  \begin{array}{ll}
\tilde{u}_{1}=\frac{\sqrt{c_1}}{\sqrt{2}\bar\gamma}\sin \sqrt{2}\bar\gamma(\eta-\eta_0)\\
\tilde{u}_{2}=\frac{\sqrt{c_1}}{\sqrt{2}\bar\gamma}\cos \sqrt{2}\bar\gamma (\eta-\eta_0)
\end{array}
 \end{eqnarray}
 as if the nonlinear dissipation does not act at all and if judged according to its solutions the equation (\ref{v-hred}) is linear.
 The other curvature cases also have corresponding nondissipative solutions.
 The only feature introduced by the reduced nonlinear Chiellini dissipation is that the amplitudes of the harmonic modes are inverse proportional to the frequency, which in fact is an Ermakov-Pinney fingerprint. Thus, one can also obtain solutions of the reduced equation (\ref{v-hred}) from the solutions (\ref{aaatheta})
 by taking ${\rm k}=0$
  \begin{eqnarray}\label{rs10b}
 \begin{array}{ll}
\tilde{u}_{-}(\eta;c_1)=\frac{\sqrt c_1}{2 |\bar\gamma|}\sqrt{-1+\cosh\big(2\sqrt 2 \bar\gamma(\eta-\eta_0)\big)}
~& \, \kappa=-1~,\\
\\
\tilde{u}_0(\eta;c_1)=\sqrt{c_1}(\eta-\eta_0)~ & \,\kappa=0~,\\
\\
\tilde{u}_{+}(\eta;c_1)=~\frac{\sqrt c_1}{2|\bar\gamma|}\sqrt{1+\sin \big(2\sqrt{2}\bar\gamma(\eta-\eta_0)\big)}
 & \, \kappa=1.
\end{array}
\end{eqnarray}
Notice also that the integration constant $c_1$ should not be zero since it occurs in the amplitude of the reduced harmonic modes.

Because of the close similarity with the undamped barotropic cosmologies and the equivalence between equations (\ref{v-heq}) and (\ref{v-meq}), we introduce the scale factors of the Chiellini barotropic universes as the roots of order $\bar\gamma$ of the $\tilde{u}$ modes, 
  \begin{eqnarray}\label{rs10c}
\begin{array}{ll}
\tilde{a}_{-}(\eta;c_1)=\left(\frac{\sqrt c_1}{\sqrt {2}|\bar\gamma|}\right)^{\frac{1}{\bar\gamma}}\left[\sinh \sqrt {2}\bar\gamma(\eta-\eta_0)\right]^{\frac{1}{\bar\gamma}}
~, & \, \kappa=-1~,\\
\\
\tilde{a}_0(\eta;c_1)={c_1}^{\frac{1}{2\bar\gamma}}(\eta-\eta_0)^{\frac{1}{\bar\gamma}}~, & \,\kappa=0~,\\
\\
\tilde{a}_{+}(\eta;c_1)=\left(\frac{\sqrt c_1}{\sqrt {2} |\bar\gamma|}\right)^{\frac{1}{\bar\gamma}}\left[\sin \sqrt {2}\bar\gamma(\eta-\eta_0)+ \cos \sqrt {2}\bar\gamma(\eta-\eta_0)\right]^{\frac{1}{\bar\gamma}} , & \, \kappa=1.
\end{array}
\end{eqnarray}

\begin{figure} [x!] 
   \centering
    \includegraphics[width= 9 cm, height=6 cm]{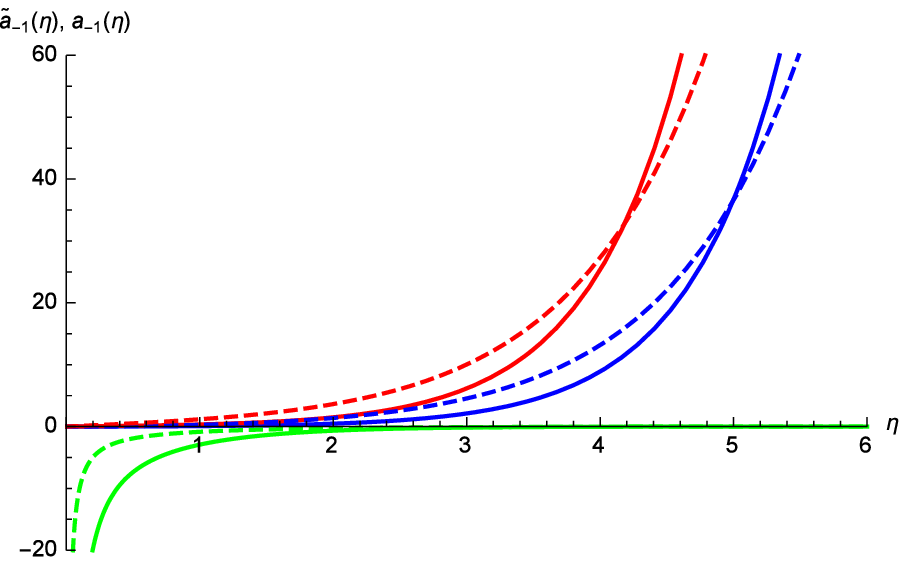}\\
    \includegraphics[width= 9 cm, height=6 cm]{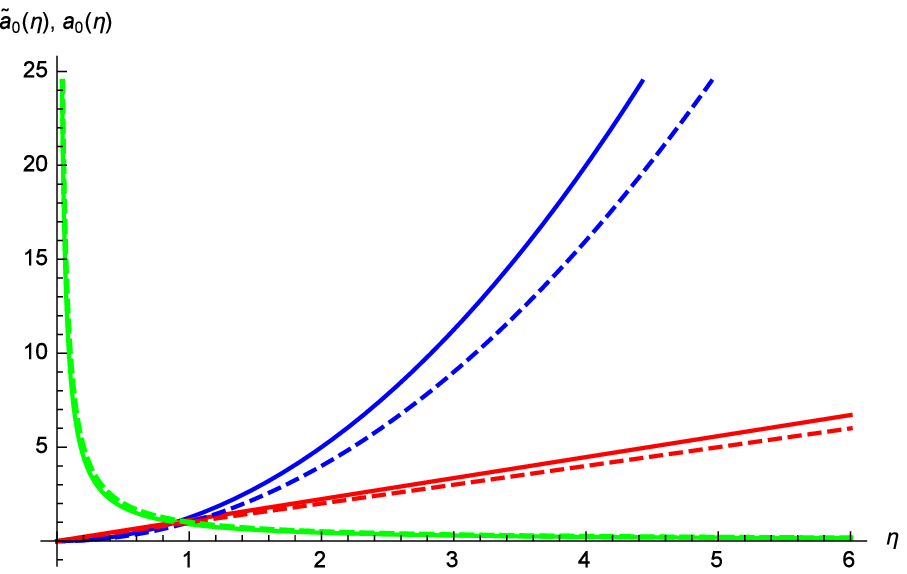}\\  
    \includegraphics[width= 9 cm, height=6 cm]{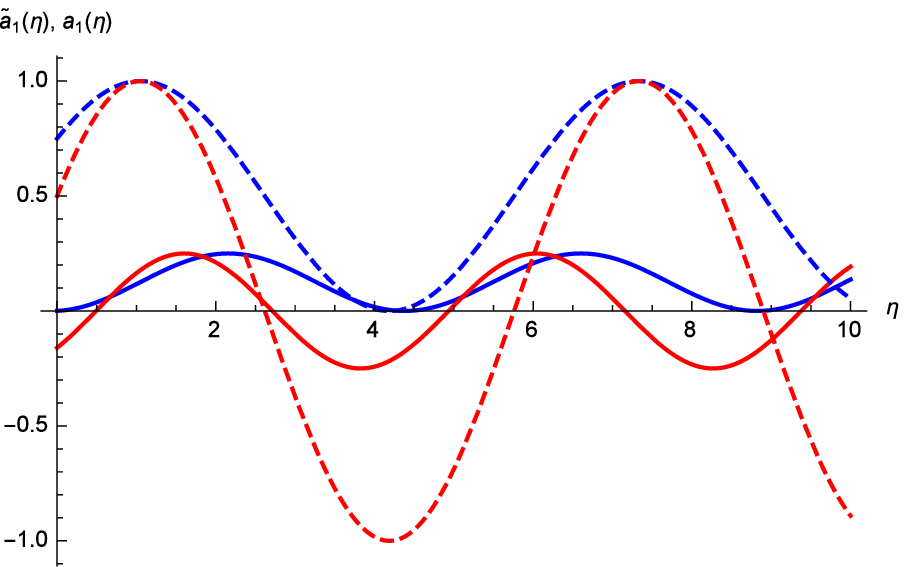}\\
    \includegraphics[width= 9 cm, height=6 cm]{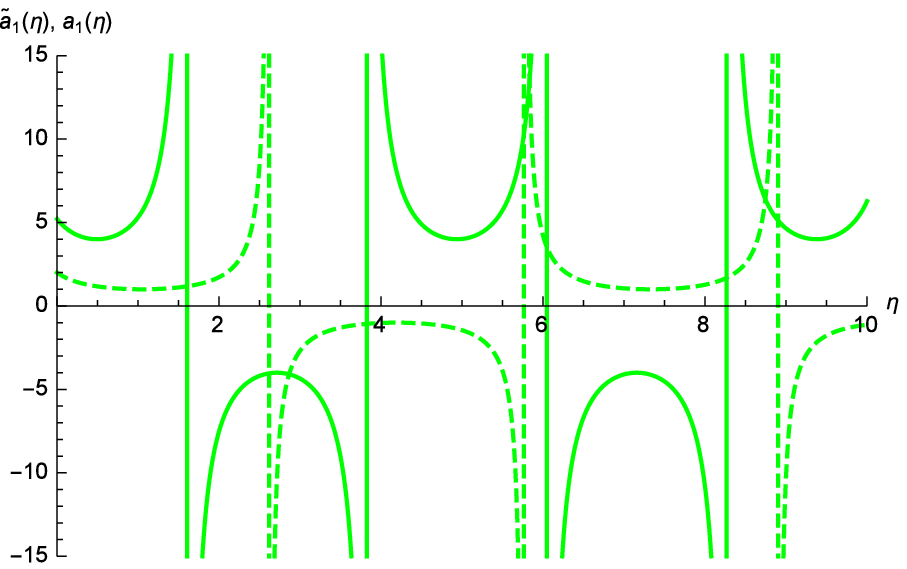}\\
\caption{\textsl{(Color online). Scale factors $ \tilde{a}_{\kappa}(\eta; c_1)$ according to equations (\ref{rs10c}) with $c_1=1/16$ for the open and closed FRW cosmologies, and $c_1=5/4$ for the flat one, respectively, (continuous lines), as compared with the standard scale factors (dashed lines) from equations (\ref{ba1}) for dust (blue), radiation (red), and vacuum (green) barotropic fluids.
}}
\label{f1}
 \end{figure}

\begin{figure} [x!] 
   \centering
    \includegraphics[width= 9 cm, height=6 cm]{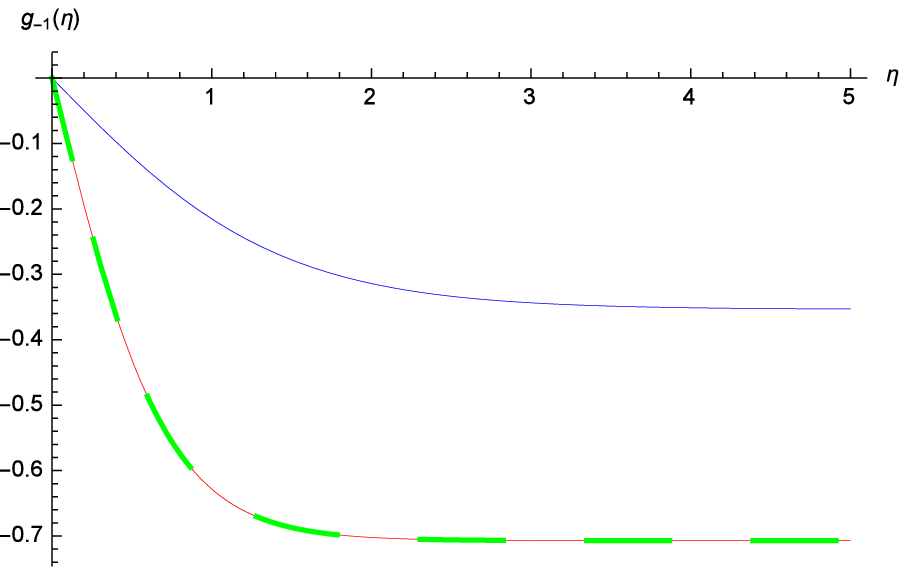}\\
    \includegraphics[width= 9 cm, height=6 cm]{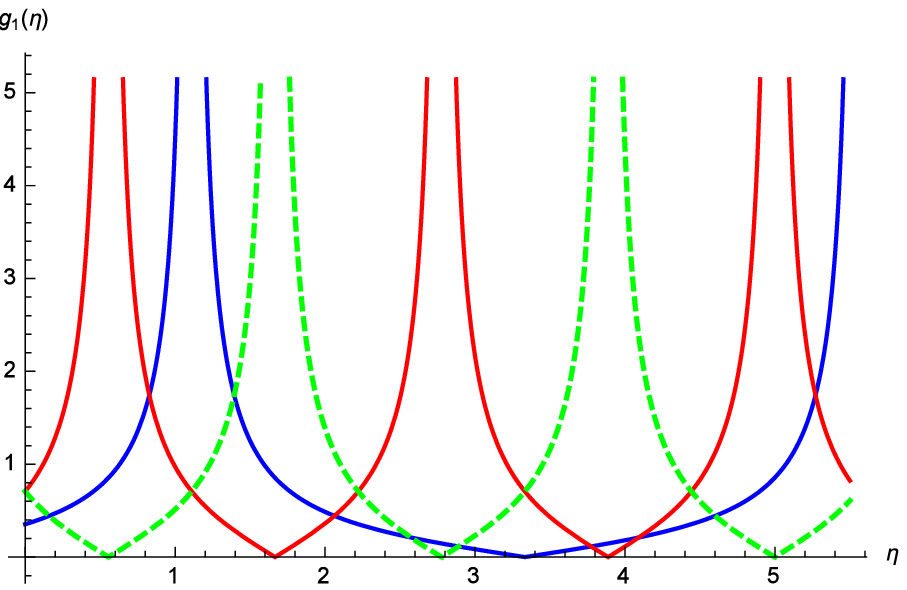}\\
\caption{\textsl{(Color online). Reduced Chiellini dissipations $g_{\kappa}(\eta; c_1)$ with $c_1=1/16$ for the open and closed FRW cosmologies, respectively, and the same barotropic fluids as in the previous figure. In the open cosmology case, the Chiellini dissipations for radiation (red continuous line) and vacuum (green long-dashed line) are identical functions of the conformal time. For the flat case, the Chiellini dissipation $g_{0}(\eta; c_1)$ is not drawn because it is always naught.}}
\label{f2}
 \end{figure}

Plots of these scale factors for dust, radiation, and vacuum  cases are presented in Fig.~\ref{f1}, and of the damping functions $g_\kappa(\tilde{u})$ in Fig.~\ref{f2}. Despite the presence of the Chiellini damping function, the scale factors of these damped barotropic universes are functionally similar to the standard scale factors of the nondissipative cosmologies. In the flat case, the reduced EP cosmology can be matched exactly to the standard cosmology by choosing $c_1=1$. In the non-flat cases, the differences occur in the amplitudes of the scale factors which are inverse proportional with the adiabatic parameter $\bar{\gamma}$ and in the rescaling of the argument of the hyperbolic and trigonometric functions. The similarity of the scale factors is due to the behavior of the Chiellini damping. As one can see in Fig.~\ref{f2}, in the case of open universes the reduced Chiellini damping is negative, i.e., it is actually a gain function, and goes rapidly to a small negative plateau. For the closed universes, the function $g_{-1}(\eta)$ may have damping regions but also periods in which it is purely imaginary.

\medskip

\begin{figure} [x!] 
   \centering
    \includegraphics[width= 9 cm, height=6 cm]{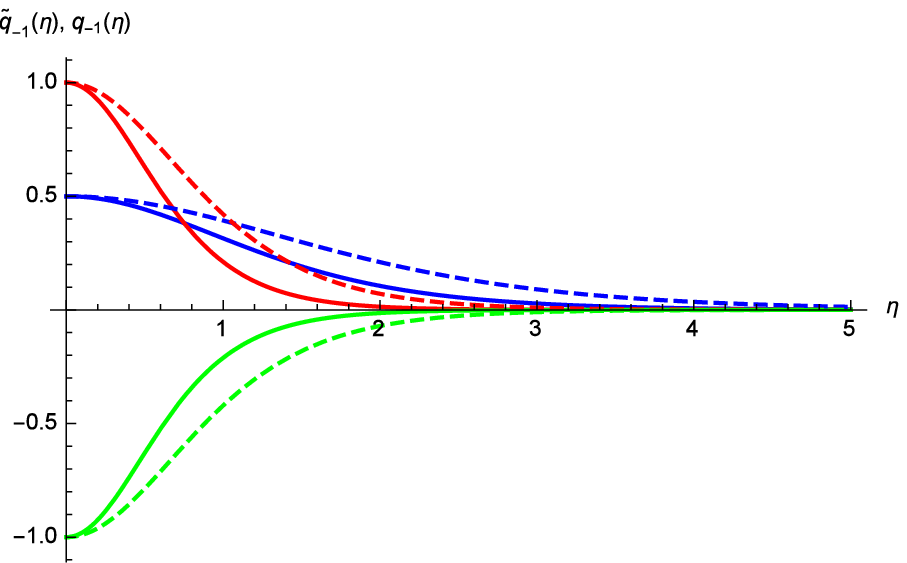}\\
    \includegraphics[width= 9 cm, height=6 cm]{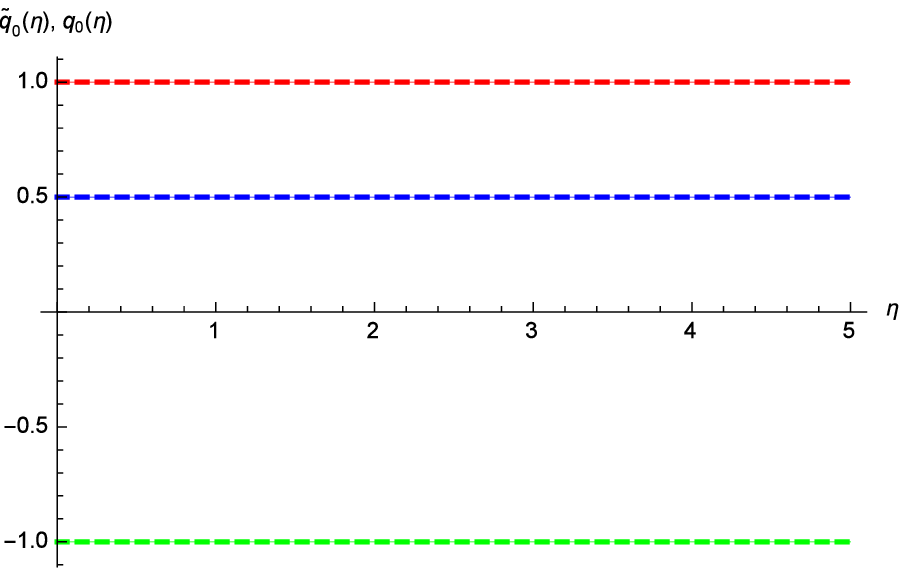}\\
    \includegraphics[width= 9 cm, height=6 cm]{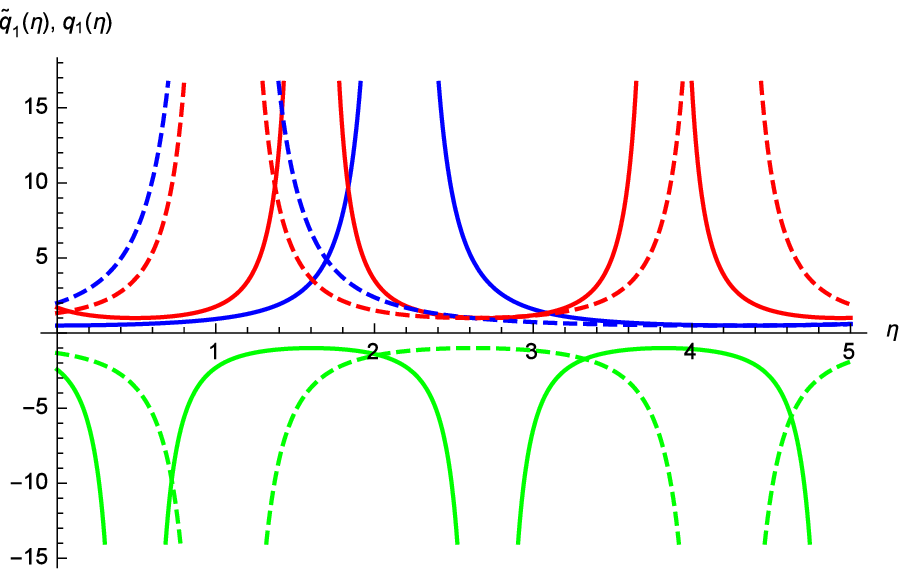}\\
\caption{\textsl{(Color online). Chiellini-damped deceleration parameters $\tilde{q}_{\kappa}(\eta; c_1)$ with $c_1=1/16$ in the reduced EP case for radiation-dominated (red, $\bar{\gamma}=1$) and matter-dominated (blue, $\bar{\gamma}=\frac 1 2$) FRW universes as given by (\ref{rs10c}). The initial phases have been chosen as naught in the first two cases and $\eta_0 =\pi/3$ in the radiation- and matter-dominated closed universes. For the flat case in the middle, the reduced deceleration parameters are identical to the standard ones.}}
\label{qtilde}
 \end{figure}

\begin{figure} [x!] 
   \centering
    \includegraphics[width= 9 cm, height=6 cm]{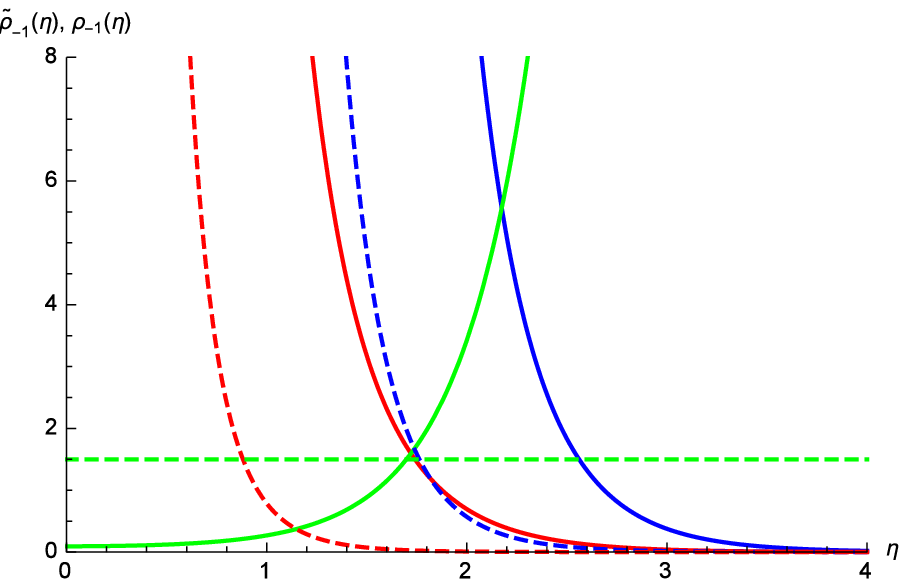}\\
    \includegraphics[width= 9 cm, height=6 cm]{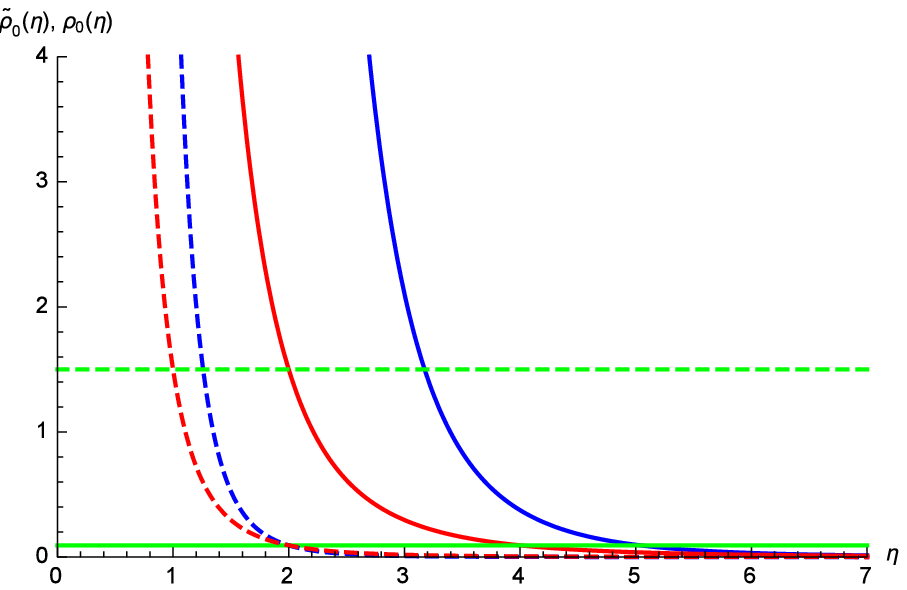}\\
    \includegraphics[width= 9 cm, height=6 cm]{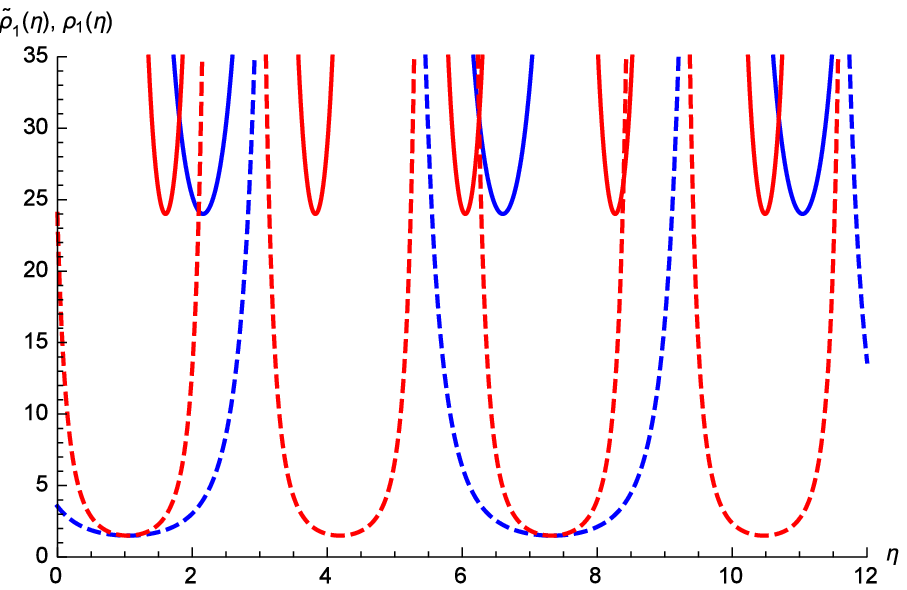}\\
  \includegraphics[width= 9 cm, height=6 cm]{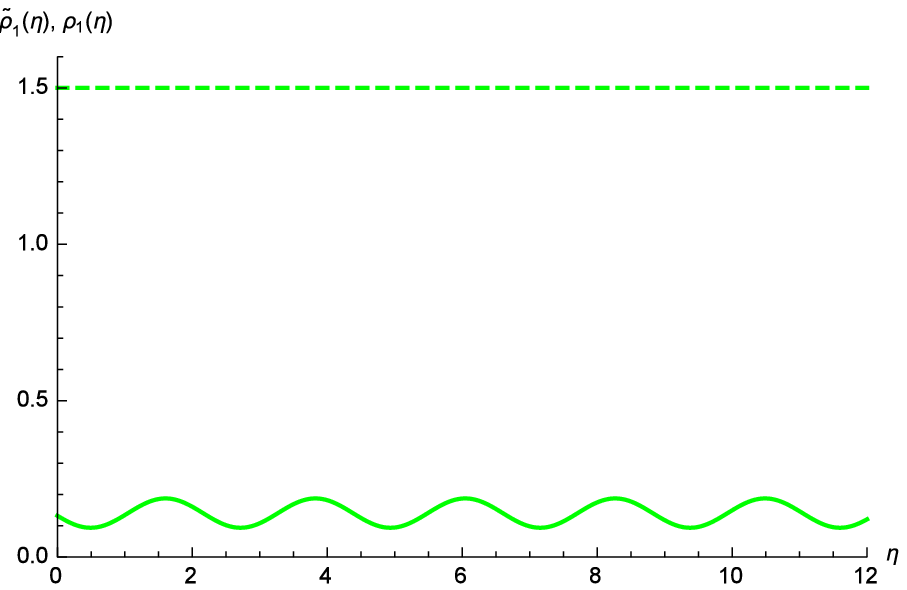}\\
\caption{\textsl{(Color online). Energy densities $\rho_{\kappa}(\eta; c_1)$ for the open, flat, and closed FRW barotropic cosmologies, from top to bottom, respectively, and the same fluids and value of $c_1$ as previously.}}
\label{fig-densities}
 \end{figure}

In Fig.~\ref{qtilde}, we plot the deceleration parameter $\tilde{q}(\eta)=1-\tilde{a}^{\prime\prime}\tilde{a}/\tilde{a}^{\prime 2}$ in the reduced EP case for the same three main barotropic cosmologies at each of the three curvature indices and compare with the standard ones. This is the basic parameter by which the occurrence of accelerating and decelerating cosmological epochs can be determined. There is no difference in the flat case with respect to the standard flat cosmology, and minor changes in shape in the non-flat cases. The energy density is plotted in Fig.~\ref{fig-densities}, according to the expression
\begin{equation}\label{enden}
\tilde{\rho}(\eta)=\frac{3}{2}\frac{\tilde{a}^{\prime 2}+\kappa \tilde{a}^2}{\tilde{a}^4}~, \qquad 4\pi G=1~,
\end{equation}
which is positive in all cases and display in general shifted values with respect to the standard cases that depend on the constant $c_1$ and the rescaled conformal time. The energy density of the reduced EP closed cosmology in the vacuum case shows an oscillatory behavior but the positivity of the energy density is maintained.

\medskip

Taking into account these results indicating that the reduced EP barotropic cosmologies are physical, we conjecture that they can be used as nonflat, either open or closed, barotropic models of the dark energy affording for accelerating late epochs because of the negative dissipation/gain of the Chiellini type. In such a framework, this dissipation is an apparent feature of the dynamics of the Universe because it is generated by the {\em nondissipative} cosmology of a barotropic fluid perceived as dark energy by the comoving observers.


\section{ Conclusion}

A class of dissipative Ermakov-Pinney equations with nonlinear dissipation of the Chiellini type is introduced in the framework of barotropic FRW cosmologies. When the nonlinear coupling constant is set to naught, the obtained damped equations provide scale factors of the non-flat universes that are similar to those of the standard barotropic cosmologies, while in the flat case, the scale factor can be made identical to the standard one by calibration. Other cosmological functions, such as the energy densities and the deceleration parameters also do not change significantly, and show that these dissipative cosmologies are viable counterparts of the standard barotropic cosmologies. Despite the almost manifest similarity, in the nonflat cases there are features that makes these cosmologies substantially different of the standard ones. One such feature is that the amplitude of the scale factors are inverse proportional to the adiabatic indices of the corresponding fluids and the other is that the functional dependence of the cosmological functions is in a scaled variable with respect to the standard barotropic cosmologies. Besides, the Chiellini dissipative function is in many cases a dissipation-gain function in the sense that, depending on the value of the parameter $c_1$, it can be negative and even purely imaginary. It should not be considered as the result of some common viscous processes which in general relativity are introduced through the techniques of the relativistic hydrodynamic formalism \cite{wm,th}. However, a physical nature can be surmised if we write it in the form $g(\tilde{u};c_1)=\bar{g}_\kappa(\tilde{u};c_1)\tilde{u}$, where $\bar{g}_\kappa(\tilde{u};c_1)=\kappa \bar\gamma^2/\sqrt{c_1-2\kappa\bar\gamma^2 \tilde{u}^2}$, which suggests a nonlinear convective origin.

\medskip

The barotropic cosmological models with rescaled adiabatic indices proposed here are based on constant equations of state (constant $\bar \gamma$) which appears to be a quite good assumption for dark energy models, because from the statistical point of view there is no evidence for a time-evolving equation of state from the entire combination of astrophysical data available at this moment \cite{dark-side}. However, even dynamical dark energy models can be accommodated, either by means of supersymmetry through which one can generate time-dependent barotropic indices \cite{robar1} or by employing simple parametrizations, such as the two-parameter Chevallier-Polarski-Linder \cite{cp,l} or recent three-parameter parametrizations \cite{novo12,hara14}.

\bigskip
\bigskip

\newpage

{\bf Acknowledgments}\\

The first two authors acknowledge Leung Center for Cosmology and Particle Astrophysics (LeCosPA) for support during their stay in Taipei.
P.C. appreciates the supports by projects from Taiwan National Science Council (NCS) and the NTU-LeCosPA. We wish to thank the referees for their useful and constructive comments and suggestions.

\bigskip
\bigskip

%

\begin{thebibliography}{999}

\bibitem{wein} S. Weinberg, {\em Gravitation and Cosmology} (Wiley, New York, 1972), pp. 475-491.
\bibitem{mis} C. W. Misner, K. S. Thorne, J. A. Wheeler, {\em Gravitation} (Freeman, San
Francisco, 1973), pp. 733-742.
\bibitem{lali} L. D. Landau, E. M. Lifschitz, {\em The Classical Theory of Fields} (Pergamon,
Oxford, 1989), pp. 363-367.

\bibitem{b93} J.D. Barrow,
Relativistic cosmology and the regularization of orbits,
The Observatory 113 (1993) 210.

\bibitem{b86} J.D. Barrow, F.J. Tipler, {\em The Anthropic Cosmological Principle} (OUP, Oxford, 1986) p. 491.

\bibitem{Far1} V. Faraoni,
Solving for the dynamics of the universe,
Am. J. Phys. 67 (1999) 732.

\bibitem{Far2} S.\c S. Bayin, F.I. Cooperstock, V. Faraoni,
A singularity-free cosmological model with a conformally coupled scalar field,
Astrophys. J. 428 (1994) 439.

\bibitem{robar1} H. C. Rosu,
Darboux class of cosmological fluids with time-dependent adiabatic indices,
Mod. Phys. Lett. A 15 (2000) 979.
\bibitem{robar2}
 H. C. Rosu, P. Ojeda-May,
Supersymmetry of FRW barotropic cosmologies,
Int. J. Theor. Phys. 45 (2006) 1191.
 \bibitem{robar3}
 H. C. Rosu, K. V. Khmelnytskaya,
Shifted Riccati procedure: Application to conformal barotropic FRW cosmologies,
SIGMA 7 (2011) 013.
\bibitem{robar4}
H. C. Rosu, K. V. Khmelnytskaya,
Inhomogeneous barotropic FRW cosmologies in conformal time,
Mod. Phys. Lett. A 28 (2013) 1340017.

\bibitem{Har-Ricc} T. Harko, F. S. N. Lobo, M. K. Mak,
A Riccati equation based approach to isotropic scalar field,
Int. J. Mod. Phys. D 23 (2014) 1450063.

\bibitem{ElN1} R.A. El-Nabulsi,
Crossing the phantom divide line from a generalized time-dependent Hubble parameter and its dynamical evolution a la Riccati,
Can. J. Phys. 91 (2013) 623.

\bibitem{ElN2} R.A. El-Nabulsi,
Nonstandard fractional exponential Lagrangians, fractional geodesic equation, complex general relativity, and discrete gravity,
Can. J. Phys. 91 (2013)  618.

\bibitem{ls09} E.V. Linder, R.J. Scherrer,
Aetherizing Lambda: Barotropic fluids as dark energy,
Phys. Rev. D 80 (2009) 023008.



\bibitem{KD} P.G. Kevrekidis, Y. Drossinos,
Nonlinearity from linearity: The EP equation revisited,
Math. Comp. Sim. 74 (2007) 196.

\bibitem{ro} H. Rosu, P. Espinoza, M. Reyes,
Ermakov approach for $Q=0$ empty FRW minisuperspace oscillators,
Nuovo Cim. B 114 (1999) 1439.

\bibitem{hl} R.M. Hawkins, J.E. Lidsey,
Ermakov-Pinney equation in scalar field cosmologies,
Phys. Rev. D 66 (2002) 023523.

\bibitem{pe} I.A. Pedrosa, C. Furtado, A. Rosas,
Exact linear invariants and quantum effects in the early universe,
Phys. Lett. B 651 (2007) 384.

\bibitem{herring} G. Herring, P.G. Kevrekidis, F. Williams, T. Christodoulakis, D.J. Frantzeskakis,
From Feshbach-resonance managed Bose-Einstein condensates to anisotropic universes:
Applications of the Ermakov-Pinney equation with time-dependent nonlinearity
Phys. Lett. A 367 (2007) 140.

\bibitem{daw} J. D'Ambroise, F.L. Williams,
A dynamic correspondence between Bose-Einstein condensates and FLRW and Bianchi I cosmology with a cosmological constant,
J. Math. Phys. 51 (2010) 062501.

\bibitem{prev} S.C. Mancas, H.C. Rosu,
Integrable dissipative nonlinear second order differential equations via factorizations and Abel equations,
Phys. Lett. A 377 (2013) 1434.

\bibitem{man} S.C. Mancas, H.C. Rosu,
Integrable Ermakov-Pinney equations with nonlinear Chiellini damping,
arXiv:1301.3567v3.

\bibitem{P} E. Pinney,
The nonlinear differential equation $\ddot{y}+p(t)y+cy^{-3}=0$,
Proc. Am. Math. Soc. 1 (1950) 681.

\bibitem{steen1874}
A. Steen, Overs. over d. K. Danske Vidensk. Selsk. Forh. 1 (1874). 
Om formen for integralet af den lineaere differentialligning af anden orden, Overs. over d. K. Danske Vidensk. Selsk. Forh. 1–12 (1874).
Translated to English by R. Redheffer and I. Redheffer, On the form of the integral of  a second-order linear differential equation,
Aequationes Math 61 (2001) 140.

\bibitem{M30} W.E. Milne,
The numerical determination of characteristic numbers,
Phys. Rev. 35 (1930) 863.

\bibitem{yano4} T. Yano, K. Kitani, H. Miyatake, M. Otsuka, S. Tomiyoshi, S. Matsushima, T. Wada, Y. Ezawa,
A high-speed method for eigenvalue problems IV. Sturm-Liouville-type differential equations,
Comput. Phys. Commun. 96 (1996) 247.

\bibitem{ran} M. Ra\~nada,
A quantum quasi-harmonic nonlinear oscillator with an isotonic term,
J. Math. Phys. 55 (2014) 082108.

%


\bibitem{wm} J. Wang, X. Meng,
Effects of new viscosity model on cosmological evolution,
Mod. Phys. Lett. A 29 (2014) 1450009.

\bibitem{th} A. Tawfik, T. Harko,
Quark-hadron phase transitions in viscous early universe,
Phys. Rev. D 85 (2012) 084032.

\bibitem{dark-side} C. Cheng, Q.-G. Huang,
Dark side of the Universe after Planck data,
Phys. Rev. D 89 (2014) 043003.

\bibitem{cp} M. Chevallier, D. Polarski,
Accelerating universes with scaling dark matter,
Int. J. Mod. Phys. D 10 (2001) 213.

\bibitem{l} E.V. Linder,
Exploring the expansion history of the Universe,
Phys. Rev. Lett. 90 (2003) 091301.

\bibitem{novo12} B. Novosyadlyj, O. Sergijenko, R. Durrer, V. Pelykh,
Do the cosmological observational data prefer phantom dark energy ?,
Phys. Rev. D 86 (2012) 083008.

\bibitem{hara14} T. Hara, R. Sakata, Y. Muromachi, Y. Itoh,
Time variation of the equation of state for dark energy,
Prog. Theor. Exp. Phys. 2014 (2014) 113E01.

\end{thebibliography}
\end{document}